# On the Fractal Geometry of the Balance Sheet and the Fractal Index of Insolvency Risk

A.K.M.**Azhar,** Vincent B.Y.**Gan,** W.A.T**. Wan-Abdullah**, H.**Zainuddin**

**Abstract**: This paper reviews the economic and theoretical foundations of insolvency risk measurement and capital adequacy rules. The proposed new measure of insolvency risk is constructed by disentangling assets, debt and equity at the micro-prudential firm level. This new risk index is the Firm Insolvency Risk Index (*FIRI*) which is symmetrical, proportional and scale invariant. We demonstrate that the balance sheet can be shown to evolve with a fractal pattern. As such we construct a fractal index that can measure the risk of assets. This index can differentiate between the similarity and dissimilarity in asset risk, and it will also possess the properties of being self-similar and invariant to firm characteristics that make up its asset composition hence invariant to all types of risk derived from assets. Self-similarity and scale invariance across the cross section allows direct comparison of degrees of risk in assets. This is by comparing the risk dissimilarity of assets. Being naturally bounded to its highest upper bound, (0,2], the fractal index is able to serve like a risk thermometer. We assign geometric probabilities of insolvency $P$ (equity is equal or less than 0 conditional on debt being greater than 0).

[1]Department of Accounting and Finance, FEM and INSPEM, Universiti Putra Malaysia, 43400 UPM Serdang, Selangor, Malaysia, akmazhar@upm.edu.my.
[2]Department of Accounting and Finance, FEM, Universiti Putra Malaysia, 43400 UPM Serdang, Selangor, Malaysia, vincentgan.upm@gmail.com.
[3]Department of Physics, Faculty of Science, University of Malaya 50603, Kuala Lumpur, Malaysia, wat@um.edu.my
[4]INSPEM, Universiti Putra Malaysia, 43400 UPM Serdang, Selangor, Malaysia, hisham@upm.edu.my

# 1. Introduction

The term "fractal" was first introduced by Mandelbrot(1967)'s seminal paper concerning the measurement of Britain's coast to describe the infinitely complex geometrical patterns that are self-similar across different scales. Fractal geometry appears most commonly in nature whether be the pattern of snowflakes or leaves (Mandelbrot 1982). However, Mandelbrot's pioneering work showed that even cotton futures (1963) and stock prices (1967) can be described by some manner of scaling which is more commonly thought to exist only in nature's fractal geometry. Subsequent research by the following generations of researchers have found that fractals exist even in stock market prices (Mandelbrot 2001; Wang et al. 2006; Wang et al. 2009), trading volumes (Moyano et al. 2006), stock market indices (Oświecimka et al. 2005; Oświecimka et al. 2006; Zunino et al. 2009), currency exchange markets (Vandewalle & Ausloos 1998), and interest rates (Cajueiro & Tabak 2007).

Fractal geometry is even used now for medical tissue imaging (Wu et al. 1991), modelling biological processes like tissue growth (Lantada et al. 2014) and the study of chromatin within the cell for cancer prognosis (Metze 2013).Nevertheless, despite these advances in the sciences, there is still a great need for the development of new fractal tools in finance. One might even go as far to say that it is an inescapable need for these tools to understand economics (Mandelbrot 2005). The purpose of this paper is to further extend the investigation of "fractality" into the balance sheet and its consequent risk of insolvency stemming from the degree of risk dissimilarity of its underlying assets. We aim to show that the balance sheet evolves within a fractal geometry and therefore insolvency risk can be represented by a fractal function as well.

# 2. Theoretical Framework

*2.1 The Balance Sheet and Insolvency Risk in an Economic Model*

*2.1.1 The Firm's Incentive*

We first begin with conventional economics characterization of asset risk and return within the balance sheets of firms and the regulator's incentive for welfare. Consider firm $i$ in an economy with $N$ firms where $N = \{i = 1, 2, 3...n\}$. Time is indexed by $t = \{0, 1\}$. The single firm $i$ choose an amount $x_i$ to invest into a single unit of risky asset. There are $Y$ units of available risky assets to finance, $Y = \{1, 2, 3...y\}$, acquiring the firm total risky assets of;

$$a_r = x_i y_i \qquad (1)$$

Risky assets, $a_r$ can be financed by debt $d_i$ and equity $e_i$. The firm chooses to use a fraction $\pi$, $\pi(d_i + e_i)$ to finance $a_r$ and stores the balance $(1-\pi)(d_i + e_i)$ into a safe asset such as cash. Thus, $a_r = x_i y_i = \pi(d_i + e_i)$. Therefore, the total assets, $a_i$ of firm $i$;

$$a_i = x_i y_i + (1-\pi)(d_i + e_i) \qquad (2)$$

For the sake of parsimony without affecting our analysis that is focused solely on book value, we do not consider asset collateralization and market based capital measures such as Value-at-risk (VaR)[1]. Total assets are naturally bounded by the accounting identity giving the budget constraint;

$$a_i = d_i + e_i \qquad (3)$$

At $t = 1$, the portfolio of assets realizes its payoff. The payoff for a single risky asset is $\check{r}$, a stochastic variable with an expected value, $E(\check{r}) = r$, $r > 0$ that is uniformly distributed over the interval $[r - z, r + z]$, with $z > 0$ as fundamental risk. The variance of the return is thus, $\sigma^2 = \frac{z^2}{3}$. Uniform density allows risk free debt contracts to be written. These may include deposits that are insured by the government. Therefore, the debt issued can be risk free if;

---

[1] VaR is the standard measure used by regulators and banks for the measurement of risk and the determination of capital adequacy. It is the bedrock capital standard in Basel II (BCBS 2004). VaR is the loss in market equity value that is a dollar quantity which is exceeded with the probability of less than or equal $\eta$: $VaR_\eta = \inf \{K \geq 0 | P(\Delta e_i \leq -K) \leq \eta\}$. A negative $K$ ensures that VaR is a positive number. Hence, VaR satisfies the condition $P(\Delta e_i \leq -VaR_\eta) \leq \eta$; the probability that losses in equity value is less than the VaR is less than or equal to $\eta$.

$$(r-z)y_i \leq d_i \tag{4}$$

A single unit of risky asset now has the market price of *p*. Assuming firms are risk averse and have mean-variance preferences, the payoff, *R*, also a random variable, of the entire asset portfolio is;

$$R = x_i y_i r + (1-\pi)(d_i + e_i) + [py_i - (d_i + e_i)] \tag{5}$$

Firms derive utility *U*, from the payoff *R*, and therefore wishes to maximize the utility function, $U = E(R) - \frac{\sigma_R^2}{\tau}$, where $\tau > 0$ is the risk tolerance of the firm and $\sigma_R^2$ is the variance of *R*, solving the following program;

$$max_{(x_i, y_i, d_i, e_i)} E(R) = x_i y_i r + (1-\pi)(d_i + e_i)$$
$$+ [py_i - (d_i + e_i)] - \frac{z^2 y_i^2}{3\tau} \tag{6}$$

Subject to constraints (3), and (4).

This formulation captures the most salient points of any firm: the fraction of assets at risk of loss, the entanglement of $d_i$, $e_i$, and $a_i$ from the balance sheet and the general risk aversion of agents.

### 2.1.2. Welfare, Monitoring, and the Regulator's Problem

In the course of enforcing regulation and monitoring the activities of banks, the regulator aims to maximize the welfare function *W*, consisting of two policy parts, $P_1$ and $P_2$; $W = P_1 + P_2$. Part 1, $P_1$ is the sum of the utilities of all banks which is reasonable to maximize the well-being of the entire system.

$$P_1 = \sum_{i=1}^{n}[E(R) - \frac{\sigma^2}{\tau}] \tag{7}$$

For the second part, the regulator considers the monitoring of aggregate debt, equity and assets in the systemic. He or she considers the fraction, $\pi$, of assets that is highly at risk of suffering catastrophic loss in value and the sufficiency of the aggregate capital cushion in the form of an equity surplus that is able to support the remaining assets. The regulator aims to minimize the loss in assets subject to the maximum threshold absorbable by the equity cushion measured by the balance between debt and equity.

$$P_2 = \min \sum_{i=1}^{n}(a_i - \pi a_i)$$
$$\text{s.t.} \sum_{i=1}^{n}(e_i - d_i) \tag{8}$$

Thus, in equilibrium $P_2 = \sum_{i=1}^{n}(a_i - \pi a_i) = \sum_{i=1}^{n}(e_i - d_i)$. Part 2 captures the aggregate threshold for adequate capital to avoid asset fire sale and credit rationing by lenders. The crux of the matter is for regulators to monitor debt and equity to control the balance between equity deficit and surplus. Thus, the regulator's problem is given as;

$$\max W = \sum_{i=1}^{n} \{[x_i y_i r + (1-\gamma)(d_i + e_i) + [py_i - (d_i + e_i)] - \frac{z^2 y_i^2}{3\tau}] + \sum_{i=1}^{n} a_i - \pi a_i\} \tag{9}$$

Hence, the major policy implication is for the regulator to choose $\pi$; the fraction of capital and assets to measure insolvency risk and institute capital adequacy standards that maximizes welfare by monitoring the firm's assets, debt, and equity.

*2.2 The Fractional Measure of Insolvency Risk*

In equilibrium, $P_2 = \sum_{i=1}^{n}(a_i - \pi a_i) = \sum_{i=1}^{n}(e_i - d_i)$. Generalizing to the individual firm and rearranging, the fractional measure of insolvency risk for the single firm *i* can be stated as;

$$\pi = \frac{a_i - (e_i - d_i)}{a_i} \tag{10}$$

We can utilize $\pi$ as a measure of insolvency risk by measuring the extent of which assets are financed by the more pernicious debt prone to sudden changes in haircuts or the safer equity normalized by total assets. It is in fact a measure of asset dissimilarity in risk. Higher values of $\pi$ mean that the fraction of assets that are higher in risk is larger. Hence, it can therefore also double up as capital adequacy rule for solvency by requiring sufficient equity to keep up proportionately with calculated values of $\pi$. The implication of expression (10) is that as $\pi$ increases, the less of assets the firm can afford to lose and more equity is required to keep it solvent.

Having provided the above economic foundations for measuring insolvency risk in the form of $\pi$, we now proceed to investigate, highlight, and provide the theoretical foundations for the geometric fractality of this new risk index.

## 3. Model Structure: The Insolvency Risk Box Approach

### 3.1 The Insolvency Risk Box (*IRBOX*) Approach

In this section, we proceed to define and elaborate the construction of our hypothesized fractional index of asset risk. Borrowing tools from the international trade literature, we proceed to construct an Insolvency Risk Box by disentangling the relationship between assets, debt, and equity following the framework of Azhar et al. (1998) and Azhar & Elliot (2006). We define the risk plane known hereafter as the Insolvency Risk Box (*IRBOX*) to be the first quadrant of the coordinate system where all coordinates of *d* and *e* are positive real numbers where, $d \in \mathbb{R}^+$ and $e \in \mathbb{R}^+$ (See Figure 1). Analysing firm *i* over *t* periods where *t* can be yearly, quarterly or monthly periods; $T = \{1, 2, 3...t\}$. By assumption, $\forall t = T$, hence, $d_t, e_t \geq 0$. The box can be extended to include firms that are already insolvent or have negative equity. However, insolvent firms are not the focus of this paper. Insolvent firms may or may not file for bankruptcy. Hence, the negative quadrant of the equity axis can be interpreted as measuring financial distress. Insolvency risk is the appropriate measure for firm risk as not all insolvent firms will file for bankruptcy but all insolvent firms will eventually restructure its balance sheet.

Alternatively, the box can be constructed from a cross section of *N* firms; *N = {i = 1, 2, 3...n}* at a specific period *t*. The dimensions of the *IRBOX* are defined as the maximum value of either *d* or *e* during the period of analysis or from the firm with the largest values respectively. Therefore, the size or area of the *IRBOX* equal the max (*d*) if $i \in t$ when the max (*d*) > max (*e*) or vice versa. Therefore, the *IRBOX* encapsulates all possible values of *d* and *e* within their natural distribution. The key innovation of the *IRBOX* is the representation of all the key elements of the balance sheet in a single geometrical representation for analysis. Thus, the area of the *IRBOX* is given by:

$$A = [max\ (d_t^2, e_t^2)], \tag{11}$$

Or $[max\ (d_i^2, e_i^2)]$.

*3.1.1 The Components of Asset Risk Similarity or Dissimilarity*

As illustrated in Figure 1a, we introduce three concepts of risk measurement from the geometric plane: the total risk (*TR*), net risk (*NR*) and the asset-capital overlap line (*ACO*). We define insolvency risk for the single firm as consisting of three components that are derived from the horizontal and vertical components of the *GEAR*, total debt and equity: Total Risk, Net Risk, and the Asset-Capital Overlap. For a detailed discussion on the construction and components of asset risk similarity, see Azhar et al. (2015).

**Definition 1.** The total risk of firm *i* is defined as: $TR = d_i + e_i$. (12)

*TR* is increasing in the North- East direction along the unity line. The locus of equi-*TR* is a line with a -1 slope. Being the sum of debt and equity, it can be interpreted that both debt and equity contribute to total insolvency risk of the firm. From the accounting identity, *TR* also represents the firm's total assets, $TR = a_i = d_i + e_i$.

Figure 1: The Insolvency Risk Box (*IRBOX*)

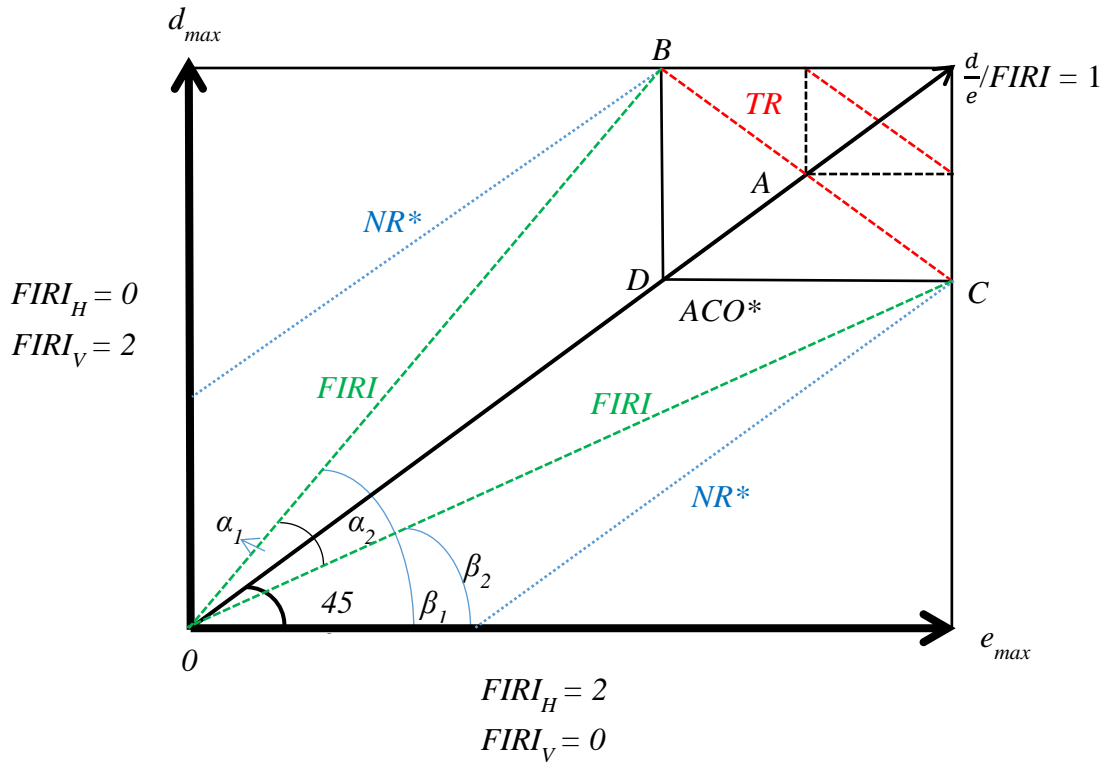

Figure 1a: Insolvency Risk Box (*IRBOX*), the *TR*, *NR*, *ACO* and *FIRI* Isoclines

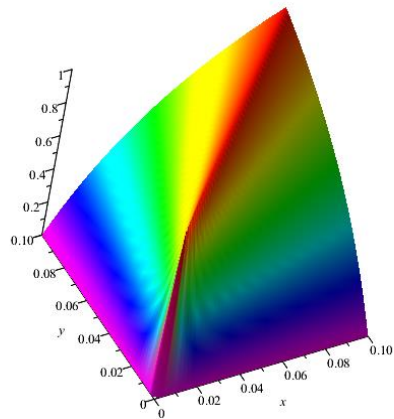

Figure 1b: *FIRI* surface plot

**Definition 2.** The net risk of firm $i$ is defined as: $NR = |d_i - e_i|$ (13)

The *NR* is the economic distance between $d$ and $e$. The locus of equi-*NR* is a has a slope of +1, is perpendicular to *TR*, increases in the North- West and South- East direction, and equals zero at the 45° line. *NR* increases as *TR* increases. *NR* measures the equity surplus or equity deficit of the firm to back the claims of its obligations with equity.

**Definition 3**. Finally, we define the asset-capital overlap, *ACO* as:

$$ACO = TR - NR = (d_i + e_i) - |d_i - e_i|$$

$$\Rightarrow ACO = a_i - |d_i - e_i| \qquad (14)$$

Thus, the *ACO* index measures the economic distance in *TR* away from the absolute balance of risk. The locus of equi-*ACO* is the L shaped line that equal zero at the origin and increases as it moves in the North- East direction with *TR*. Above the unity line, *ACO* value will consist of values of $d_i > e_i$, and $e_i > d_i$ below the unity line. Values of $d_i > e_i$ will be on the vertical *ACO* line and values of $e_i > d_i$ will be on the horizontal *ACO* line. This index measures the degree of similarity of which total assets are financed by equity surplus or equity deficit.

*3.1.2 The Fractional Measure of Insolvency Risk: Measuring Asset Risk Similarity*

We now characterize a new measure of total risk posed by the interplay of debt and equity that characterizes how total risk increases with net risk. This relationship is captured by the asset-capital overlap line (*ACO*) as *TR= ACO + NR*.

Therefore, we ask the question, how much of risk shared by debt and equity as represented by the *ACO* contributes to the total risk (*TR*) of the firm. In other words, how similar in risk are the assets for a given level of debt and equity. Hence, we propose the fraction of *ACO* in *TR* as:

**Definition 4.** Fraction of asset-capital overlap, $FACO = \frac{ACO}{TR}$ \qquad (15)

This ratio is symmetric, proportional and scale invariant as it increases proportionally with the increase in total risk and therefore is a better representation of risk. Figure 1a illustrates the operation of this principle. Therefore, we now define, the *FACO* now known hereafter as the Firm Insolvency Risk Index (*FIRI*).

**Definition 5**. $FIRI = \frac{ACO}{TR} = \frac{(d_i+e_i) - |d_i-e_i|}{(d_i+e_i)} = 1 - \frac{|d_i-e_i|}{(d_i+e_i)}$

$$\Rightarrow FIRI = 1 - \frac{|d_i-e_i|}{(a_i)} \quad (0,2] \quad (16)$$

The expression in (16) can be decomposed to its horizontal, $FIRI_h$, and the vertical, $FIRI_v$, components given by:

$$FIRI_h = 1 - \frac{d_i-e_i}{(d_i+e_i)} \quad (0,2] \quad (17)$$

$$FIRI_v = 1 + \frac{d_i-e_i}{(d_i+e_i)} \quad (0,2] \quad (18)$$

By removing the modulus sign, we arrive at both $FIRI_h$ and $FIRI_v$ which give us measures of risk dissimilarity. $FIRI_h$ measures the degree of dissimilarity which assets are financed by equity and $FIRI_v$ measures the degree of asset dissimilarity in terms of being financed by debt. Interpretation of risk is easy and comparable across all firms. For $FIRI_h$ smaller values signify higher risk and for $FIRI_v$ higher values mean higher risk. Figure 1b shows the *FIRI* risk surface. Thus in Figure 1a, as the angles of $\alpha_1$ and $\alpha_2$ are equal, every point of coordinates on the two equi-*FACO* rays share an equal *FACO* value that accounts for the horizontal and vertical contribution to total asset risk. The locus of equi-*FACO* consists of all points and only those points whose *d* and *e* coordinates share a common $\frac{(d_i+e_i) - |d_i-e_i|}{(d_i+e_i)}$ value. As illustrated in Figure 1a, the *FIRI* ray deceases towards a slope of unity as it sweeps from the vertical axis (*d*) towards the *GEAR* = 1 line and increases towards a slope of unity as it sweeps from the horizontal axis (*e*) towards the *GEAR* = 1 line.

## 4. Fractal Analysis of the Balance Sheet

In the course of this paper, geometrical hints of the 'fractured' nature of insolvency risk and the balance sheet were presented repeatedly. In fact, it is by no coincidence that the *FIRI* index is a fractional measure of risk. We now consider the geometry of the 'fractured' nature of insolvency risk and the balance sheet as presented above. Ultimately, we intend to show that the balance sheet evolves across time and cross section through a fractal geometry that can be captured with the appropriate functional form that represents the riskiness of its assets and the safety of its equity. Figure 1 shows how the *FIRI* line scales proportionately with the total assets line sweeping across its resting on the vertical and horizontal axes before settling on the balanced risk line where $d/e= 1$. Clearly seen in Figure 1 is also how the *ACO* and the total assets lines form a right triangle that scales in size until its convergence into the north east corner of the *IRBOX*. This suggests that self- similarity is present.

A fractal dimension is defined as a statistical index of complexity that compares how changes in details change with the scale which it is measured (Falconer 2003). Mandelbrot first introduced the term 'fractal' to describe these patterns or sets. Unlike topological dimensions that have integer values, fractals have dimensions that are not integers but rather 'in between'(Mandelbrot, 1967). Thus, while many day-to-day objects have a smooth dimension, fractals display a 'rough' geometry. For instance, the shapes that people are most familiar with like the circle, square, triangle or rectangle has smooth geometrical dimension versus shapes in nature such as snowflakes or fern leaves. We proceed to provide a simple analysis of the fractality of the *IRBOX*. We believe this is the first attempt to suggest that the balance sheet displays the properties of fractal geometry and it is this property that allows us to design the *FIRI* index which is symmetrical, scale invariant and proportional.

Figure 2. The geometric progression of the *TR*, *NR*, *ACO*, and *FIRI* isoclines within the *IRBOX*.

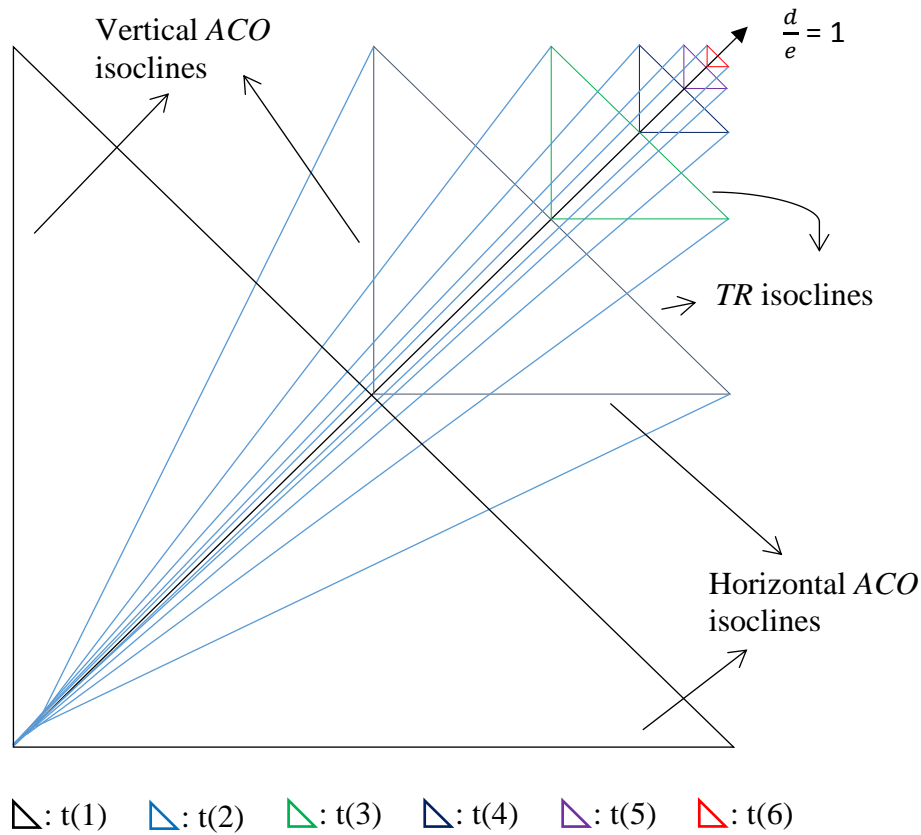

△ : t(1)   △ : t(2)   △ : t(3)   △ : t(4)   △ : t(5)   △ : t(6)

For illustrative purpose we consider Figure 2 where the *IRBOX* from Figure 1 is decomposed to six right triangles. Each right triangle is composed by the *TR* or total assets line as the hypotenuse and the *ACO* lines as sides. Figure 2 clearly shows how the *FIRI* line scales proportionately with the length of the *TR* line, the area and perimeter of each successive iterated triangle from triangle 1 denoted as t(1) to triangle 6, t(6). The *ACO* isoclines also scale with the *TR* line, the area and perimeter of each triangle. The *NR* line scales similarly as the *FIRI* line although not shown in the figure. Each successive triangle is a continuous contraction map of the first and is a subset of the *IRBOX*. Triangles t(2) to t(6) are therefore similitudes of t(1). As the area of the triangles converges to zero, the *FIRI* line converges unto the $d/e = 1$ line. Therefore, t(1) is invariant to each iteration. This implies that the *FIRI* line at rest on the horizontal and vertical axes of t(1) is also invariant to changes in *TR* which means that *FIRI* scales proportionately with *TR*. Thus, the changes in the detail of the *FIRI* line changes with the scale it is measured with. The figure implies that for every coordinate (*x*, *y*), there exist a common *NR*, *ACO* and *FIRI* value that scales proportionately with *TR*. Figure B1 also shows an interesting geometry where the *d/ e* = 1 line is the orthonormal basis of the *IRBOX*. Hence, this implies that each *NR* line is also the orthonormal basis for every right triangle bounded within the *IRBOX*.

*4.1 The Fractal Dimensions of the IRBOX: The Origins of Symmetry, Scale Invariance and Proportionality.*

The fractal dimension is given as, $D = \frac{\log N}{\log e}$, where *N* is the number of new shapes or boxes and *e* the degree of magnification. This formulation is known as the box counting method to determining the fractal dimension and is the easiest to apply (Falconer, 2003). Consider Figure 3 below which shows a regular unit square being magnified two times and producing four new squares. Thus, the dimensions of the square; $\frac{\log 4}{\log 2} = 2$, a whole integer. Whole integer dimensions are common to most man made geometry.

Figure 3: The dimensions of a square

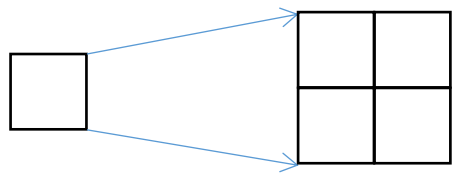

We now consider Figure 4 where the *IRBOX* is represented as a unit square in five iterations. As the *IRBOX* is a perfect unit square, it is actually composed of two right triangles in s(1). Thus, there exists symmetry on both diagonals. In the next step, s(2), we remove two triangles. As explained earlier, the *TR* and *ACO* lines form a right isosceles triangle. Therefore, focusing on each right isosceles triangle, we have two right angled Sierpinski Triangles. Each triangle removed is two times smaller than original in s(1) and three new triangles are produced. Alternatively, we can view the problem by stating that the iteration s(2) as the first square removed which is two times smaller than the original of s(1). This would leave us with three new squares. Thus, we have, *e* = 2, and *N*= 3

**Theorem 1**. The fractal dimension of the *IRBOX* Gasket is $\frac{\log 3}{\log 2}$ = 1.585.

This result is simple and it is clear that the fractal dimension of the IRBOX Gasket is the same as the Sierpinski Gasket. Proceeding with five iterations, we arrive at the fifth iterated set, s(5) which we denote as the *IRBOX* Gasket. We stop at five iterations for the sake of parsimony. In theory, the iterations can proceed to approach infinity and the triangles fill up the risk space within the *IRBOX* implying that the *IRBOX* has a finite area but infinite perimeter. We thus provide graphical illustration on how the fractional geometry of the balance sheet looks like in s(5).

Figure 4: The *IRBOX* and five successive iterations

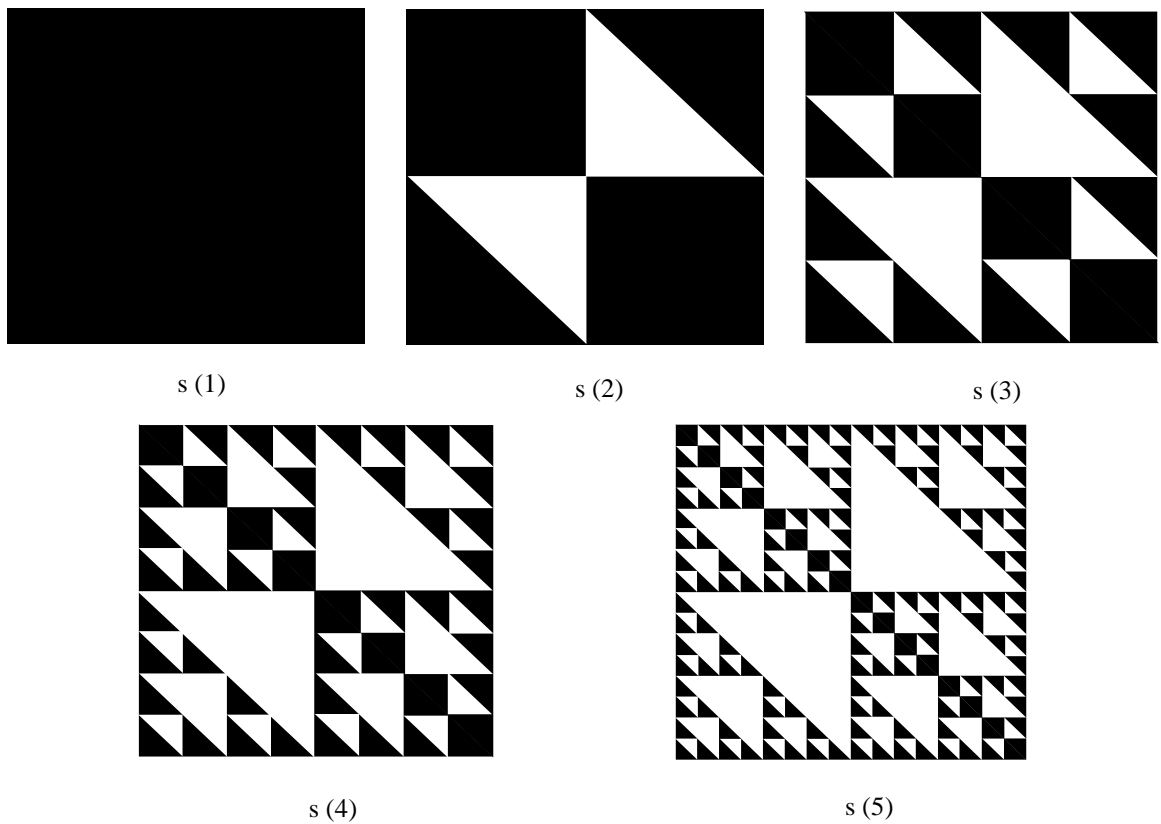

s (1)    s (2)    s (3)

s (4)    s (5)

**Theorem 2**. The *IRBOX* Gasket has an area of 1 and an infinite perimeter.

**Proof** (Alternate proof of encapsulation). With each iteration, the number of triangles removed follows a geometric progression where beginning with 2 in s(2), we proceed to 6 in s(3) and 18 in s(4). Thus, the number of triangles removed with each $k$ iteration: 2, 2(3), $2(3^2)...2(3^{k-1})$. The length of each side follows by 1/2, 1/4, 1/8...$1/2^k$. The area removed with each iteration is therefore $2(3)^{k-1}(1/2^k)$. Thus, the total area removed after $k$ iteration follows a geometric progression $2[1/4 + 3/16 + 9/64 +....2(3)^{k-1}(1/2^k)]$. The sum of a geometric sequence is given by $S_k = a_1 \frac{1-r^k}{1-r}$, where $a_1$ is the first term and $r^k$ the common ratio. In this case it is apparent that $a_1 = 1/4$ and $r^k = 3/4$. Thus, the total area removed after $k$ iterations is $S_k = (1/4)\frac{1-3/4^k}{1-3/4} = 1 - (3/4)^k$. The limiting value of the total area removed as $k$ approaches infinity is $S_\infty = \lim_{n\to\infty} 1 - 3/4^k = 1$.

Let us now consider the perimeter calculations. The first iteration in s(1) shows that when two triangles are removed, six triangles remain. Thus, after $k$ iterations, the number of triangles remaining would be $2(3)^k$. The perimeter of each remaining triangle would be $\frac{2+\sqrt{2}}{2^k}$ after the $k$-th iteration. Thus, the total perimeter of all remaining triangles after the $k$-th iteration would be $2(3)^k \frac{2+\sqrt{2}}{2^k}$. Let X be $2 + \sqrt{2}$. This leads us to a geometric sequence of $2X(3/2 + 9/4 + 27/8 +....\frac{3^k}{2^k})$. It is now apparent that $a_1 = 3/2$ and $r^k = 3/2$. Thus, the total perimeter remaining after $k$ iterations is $S_k = (3/2)\frac{3/2^k-1}{3/2-1} = 3[(3/2)^k - 1]$. The limiting value of the total perimeter remaining as $k$ approaches infinity is $S_\infty = \lim_{n\to\infty} 3[(\frac{3}{2})^k - 1] = \infty$. This completes the proof. This is also an alternate proof for encapsulation as an infinite perimeter would mean that every infinitesimal point would be captured within the box.

In Figure 5, we can fully visualize the *TR*, *NR*, *ACO* and *FIRI* isoclines holistically within the *IRBOX*. Here we can see that each right isosceles triangle represent a *TR* and *ACO* line and the triangles are self- similar and fill up the entire space of the box. This property of self- similarity allows the *FIRI* index to scale proportionately with total assets that represent the size of the balance sheet. Thus, the *FIRI* index can change along with the scale that it is being measured at with the triangles in red attempting to convey this message of scaling. This implies that by applying the *IRBOX* and using the *FIRI* index, no distributional assumptions such as i.i.d or normality is required.

Figure 5: The Fractal Geometry of the *IRBOX*

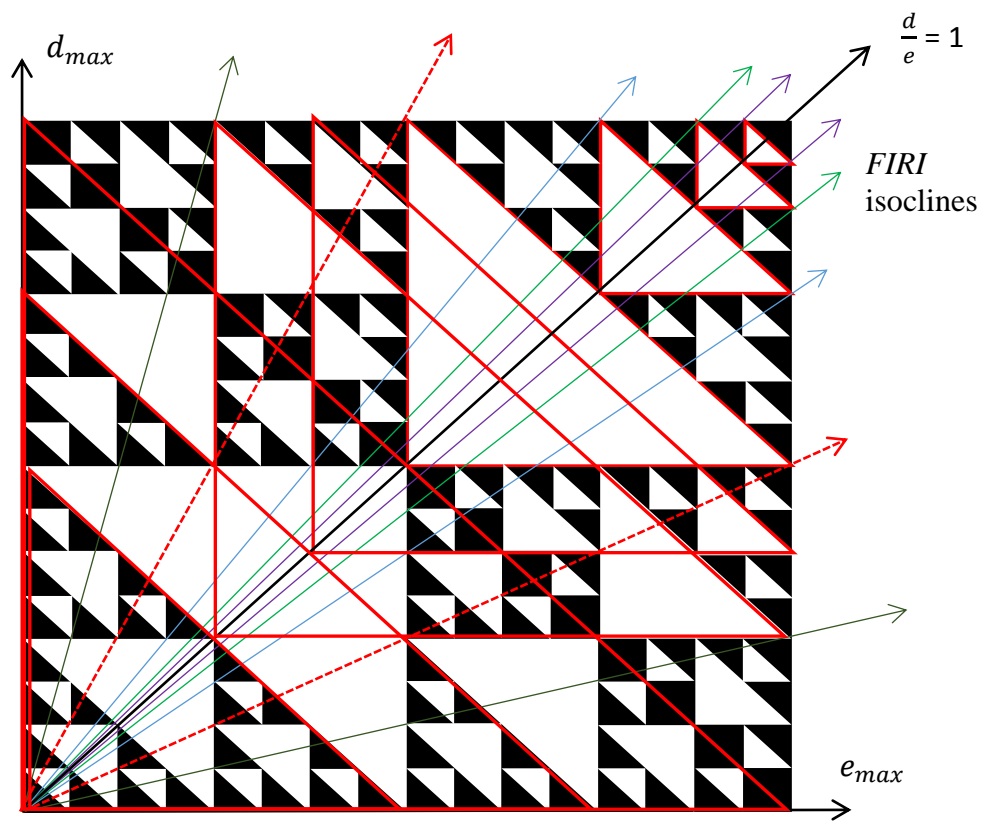

## 5. Conclusions

Fractal geometry has long been suggested by Mandelbrot to be an indispensable tool in finance and it is no accident that such tool appears in the discussion of the geometry of the balance sheet. The scaling symmetry origin of fractals can also be found in the various fields of finance and economics exhibited by prices and other market behavior. The purpose of this paper is to show that with the appropriate function, the balance sheet can be shown to evolve with a fractal pattern. Thus, self- similarity is an inherent property of the balance sheet. Intra-industry trade (IIT) would therefore be governed by the similar fractal geometry discussed here as our *FIRI* index is similar in functional form to the Grubel-Lloyd (Grubel & Lloyd 1971) index for IIT measurement. Since the balance sheet can be shown to evolve in a fractal pattern, we can construct a fractal index that can measure the risk of assets. The index can differentiate between the similarity and dissimilarity in asset risk. Furthermore, the constructed index is a fractal index, it will possess the properties of being self-similar and invariant to firm characteristics. These are characteristics that make up its asset composition. Thus, invariant to all types of risk derived from assets. Self-similarity and scale invariance across the cross section allows direct comparison of degree of risk in assets. Being naturally bounded to its highest upper bound, (0,2], the fractal index is able to serve like a risk thermometer. We can further assign geometric probabilities of insolvency *P* (equity is equal or less than 0 conditional on debt being greater than 0).